\documentclass[aps,prb,showpacs]{revtex4}
\usepackage{mathrsfs}
\usepackage{amssymb}
\usepackage{amsmath}
\usepackage{bm}
\usepackage{graphics}
\usepackage{natbib}

\begin{document}
\title{Josephson effects in the junction formed by \emph{DIII}-class topological and $s$-wave superconductors with an embedded quantum dot}

\author{Zhen Gao} 
\author{Wan-Fei Shan}
\author{Wei-Jiang Gong}\email{gwj@mail.neu.edu.cn}

\affiliation{1.
College of Sciences, Northeastern University, Shenyang 110819, China}
\date{\today}

\begin{abstract}
We investigate the Josephson effects in the junction formed by \emph{DIII}-class topological
and $s$-wave superconductors, by embedding a
quantum dot in the junction. Three dot-superconductor coupling
manners are considered, respectively. As a result, the Josephson current is found to oscillate in $2\pi$ period. Moreover, the presence of Majorana doublet in the \emph{DIII}-class superconductor renders the current finite at the case of zero phase difference, with its sign determined by the fermion parity of such a jucntion. In addition, the dot-superconductor coupling plays a nontrivial role in adjusting the Josephson current. When the $s$-wave superconductor couples to the dot in the weak limit, the current direction will have an opportunity to reverse. The results in this work will be helpful for understanding the transport properties of the \emph{DIII}-class superconductor.
\end{abstract}

\maketitle

\section{Introduction}
Topological superconductors (TSs) have received much attentions from
experimental and theoretical aspects because Majorana modes appear
at the ends of the one-dimensional TS which can potentially be used
for topological quantum computation.\cite{Majorana1,RMP1,Zhang2} Due
to the possibility of achieving Majorana modes, the systems with TSs
show abundant and interesting physical
characteristics.\cite{Majorana2} For instance, in the
proximity-coupled semiconductor-TS devices, the Majorana zero modes
induce the zero-bias anomaly.\cite{Fuliang} A more compelling TS
signature is the unusual Josephson current-phase relation. Namely,
when the normal s-wave superconductor nano-wire is replaced by a TS
wire with Majorana zero modes, the current-phase relation will be
modified to be $I_J \sim {\cal P}\sin{\phi\over2}$ with the period $4\pi$ ($\phi$ is the
superconducting phase difference and $\cal P$ is the fermion parity). This is the so-called the
fractional Josephson
effect.\cite{Josephson1,Josephson2,Josephson3,Aguado}
\par
More recently, the time-reversal invariant TSs, i.e., the \emph{DIII}
symmetry-class TSs,\cite{Ryu,Qi,Teo,Timm,Beenakker2} have become one new
concern.\cite{Deng,Nakosai,Wong} In such a kind of TSs, the zero
modes appear in pairs due to Kramers's theorem, differently from the
chiral TSs. Consequently, two Majorana bound states (MBSs) will be
localized at each end of the \emph{DIII}-TS nanowire, leading to the
formation of one Kramers doublet.\cite{Zhang,Nagaosa2} Since the
Kramers doublet is protected by the time-reversal symmetry, it will
inevitably drive some new transport phenomena, opposite to the
single Majorana zero mode. Up to now, many groups have proposed
proposals to achieve the \emph{DIII}-TS nanowires, by using the
proximity effects of $d$-wave, $p$-wave, $s\pm$-wave, or
conventional $s$-wave
superconductors.\cite{Stern,Flensberg,Haim,ZhangF2,Sau,Loss}
Meanwhile, physicists have paid attention to the quantum transport
phenomena contributed by the Kramers doublet, and some important
results have been reported.\cite{Qixl,Liuxj,Gong1} For instance, in the
Josephson junction formed by the Majorana doublet, the period of the
Josephson current will be varied by the change of fermion parity in
this system.\cite{Liuxj} This exactly embodies the nontrivial role
of Majorana doublet in contributing to the Josephson effect.
However, for further understanding the property of Majorana doublet, 
Josephson effects in any new junctions should be investigated.
\par
In this work, we would like to discuss the Josephson effect in a complicate
junction, i.e., one junction formed by the indirect coupling between
a \emph{DIII}-class TS and $s$-wave superconductor with one embedded quantum dot
(QD). Our calculations show that the interplay
between the \emph{DIII}-class TS and the $s$-wave superconductor induce
interesting results. Although the Josephson current oscillates in $2\pi$ period, the presence of Majorana doublet in the \emph{DIII}-class TS renders it finite at the case of zero phase difference, with its sign determined by the fermion parity of the whole system. Also, QD plays a nontrivial role in adjusting the Josephson effect. To be concrete, in the extreme case where the $s$-wave superconductor couples weakly to the
QD, the direction of the Josephson current will be reversed. All these results embodies the specific role of \emph{DIII}-class TS in driving the Josephson effect.

\begin{figure}
\begin{center}
\scalebox{0.54}{\includegraphics{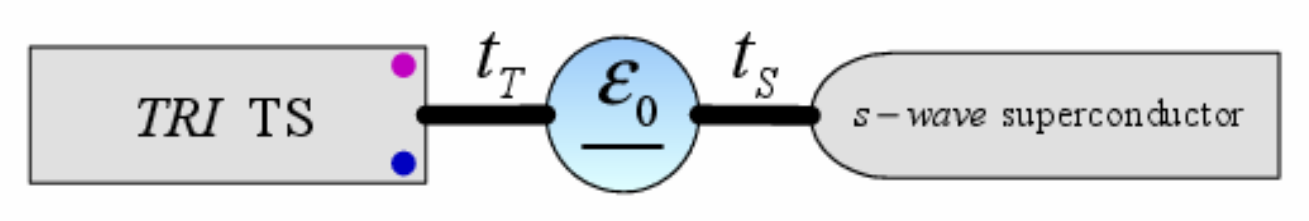}} \caption{The junction
formed by \emph{DIII}-class topological and $s$-wave superconductors. One
QD is embedded in the junction.} \label{Struct}
\end{center}
\end{figure}

\section{Theoretical model}
The considered Josephson junction is illustrated in
Fig.\ref{Struct}, where one \emph{DIII}-class TS couples to one $s$-wave
superconductor via one QD. The Hamiltonian of such a system can be
written as $H=H_{p}+H_s+H_D+H_T$. The first two terms, i.e., $H_p$
and $H_s$, denote the Hamiltonians of the \emph{DIII}-class TS and s-wave
superconductor respectively, which is written as\cite{Qixl}
\begin{eqnarray}
H_p&=&-\mu_p\sum_{j\sigma}c_{j\sigma}^\dagger c_{j\sigma}-
t\sum_{j\sigma}(c_{j+1,\sigma}^\dagger c_{j\sigma}+\text{h.c.})+\sum_{j\sigma\sigma'}
[(-i\sigma_1\sigma_2)_{\sigma\sigma'}\Delta_p c_{j+1,\sigma}^\dag c_{j\sigma}^\dagger+\text{h.c.}],\notag\\
H_s&=&-\mu_s\sum_{j\sigma}f_{j\sigma}^\dagger
f_{j\sigma}-t'\sum_{j\sigma}(f_{j+1,\sigma}^\dagger
f_{j\sigma}+\text{h.c.})+\sum_j(\Delta_s f_{j\uparrow}^\dagger
f_{j\downarrow}^\dagger+\text{h.c.}).
\end{eqnarray}
$c^\dag_{j\sigma}$ and $f^\dag_{j\sigma}$ ($c_{j\sigma}$ and $f_{j\sigma}$) are the
electron creation (annihilation) operators in the \emph{DIII}-class TS and $s$-wave
superconductor, respectively. $\mu_p$ and $\mu_s$ are the chemical
potentials of the superconductors, and $\Delta_p$ and $\Delta_s$ are
the Copper-pair hopping terms. $H_D$ describes the Hamiltonian of
the QD. For a single-level QD, it takes the form as $H_D=H_{d0}+H_{ee}$ with
\begin{eqnarray}
H_{d0}=\varepsilon_0\sum_\sigma(n_\sigma-{1\over2}), ~H_{ee}=U(n_\uparrow-{1\over2})(n_\downarrow-{1\over2}).
\end{eqnarray}
$n_\sigma=d^\dag_\sigma
d_\sigma$ is the electron-number operator, in which $d^\dag_\sigma$ and $d_\sigma$ are the creation and annihilation
operators in the QD. $\varepsilon_0$ is the QD level, and $U$ denotes the intradot Coulomb repulsion. $H_T$, the last term of $H$,
represents the couplings between the QD and the superconductors. It
can be given by
\begin{eqnarray}
H_T&=-t_T\sum_\sigma(e^{i\theta_T/2}c_{1\sigma}^\dag
d_\sigma+\text{h.c.})-t_S\sum_\sigma(e^{i\theta_S/2}f_{1\sigma}^\dag d_\sigma
+\text{h.c.}),
\end{eqnarray}
where $t_T$ and $t_S$ are the QD-superconductor coupling
coefficients, respectively.
\par
It is well-known that the phase difference between superconductors drives finite current through
one Josephson junction. With respect to such a junction, the current properties can be evaluated by the following formula
\begin{equation}
I_J={2e\over \hbar}{\partial {\langle H}\rangle\over\partial
\theta}.
\end{equation}
In this equation $\theta=\theta_T-\theta_S$ is the phase difference between the
superconductors, and $\langle \cdots\rangle$ is the thermal average.
As a typical case, i.e., the zero-temperature limit, the Josephson
current can be simplified as
\begin{equation}
I_J={2e\over\hbar}{\partial E_{GS}\over\partial \theta}.
\end{equation}
This formula shows that the calculation about the Josephson
current is dependent on the ground-state (GS) level of this system.
\par
In the low-energy region, \emph{DIII}-class TS only contributes Majorana doublets to the Josephson effect, so $H_p$ describes the coupling
between two Majorana doublets. For the extreme case of one
infinitely-long TS, the coupling strength between the Majorana
doublets decreases to zero.\cite{LossN} Following this idea, we
project $H_T$ onto its zero-energy subspace. As a result, $H$ can be rewritten
as
\begin{eqnarray}
H&=&\varepsilon_0\sum_\sigma (n_\sigma-{1\over2})+U(n_\uparrow-{1\over2})(n_\downarrow-{1\over2})+\sum_{k\sigma}\xi_kf^\dag_{k\sigma}f_{k\sigma}+\Delta_s\sum_kf^\dag_{k\uparrow}f^\dag_{-k\downarrow}
\notag\\
&&+t_T\sum_\sigma(e^{i\theta/2}d_\sigma-e^{-i\theta/2}d_\sigma^\dag)\gamma_{0\sigma}-\sum_{k\sigma}V_{kS}
f_{k\sigma}^\dag d_\sigma+\text{h.c.},
\end{eqnarray}
where $H_s$ has been projected into in the Bloch space with $\xi_k=-\mu'-2t'\cos k$ and $V_{kS}={1\over \sqrt{N}}\sum_k
e^{ik}t_S$. $\gamma_{0\sigma}$, the Majorana operator, which obeys
the anti-commutation relationship of
$\{\gamma_{0\sigma},\gamma_{0\sigma'}\}=2\delta_{\sigma\sigma'}$.
Based on the renewed expression of $H$, we next try to diagonalize
the Hamiltonian of such a Josephson junction.

\section{Diagonalization of the junction Hamiltonian}
The continuum state in the $s$-wave superconductor hinders the
diagonalization of the system's Hamiltonian. In order to present a comprehensive analysis, we would like to consider three cases, i.e., the cases of $t_T\gg t_S$, $t_T\approx  t_S$, and $t_T\ll t_S$, followed by the utilization of different approximation methods. The following are the detailed
discussion processes. For convenience, they are named as Case I,
Case II, and Case III, respectively.
\subsection{Diagonalization of $H$ in Case I}
In Case I where $t_T\gg t_S$, the subsystem formed by the QD and Majorana doublet can be considered to be one system, whereas the $s$-wave
superconductor can be viewed as a perturbation factor. We next simplify the system Hamiltonian by means of the perturbation theory. Ignoring the Coulomb interaction term in the QD, we can write out the action of the subsystem of QD and $s$-wave superconductor
\begin{small}
\begin{eqnarray}
S&=&\int d\tau\,[\Psi_d^\dagger(\partial_\tau+\mathcal{H}_{d0})\Psi_d]+\int d\tau\sum_k\{\Psi_k^\dagger[\partial_\tau+\mathcal{H}_s(k)]\Psi_k-[\Psi_d^\dagger {\cal V}_{kS}\Psi_k+\Psi_k^\dagger {\cal V}^\dag_{kS}\Psi_d]\},
\end{eqnarray}
\end{small}
where ${\cal H}_{d0}=\varepsilon_0\sigma_z$, ${\cal H}_s=\xi_k\sigma_z+\Delta_s\sigma_x$, and ${\cal V}_{kS}=V_{kS}\sigma_z$. As for the field operators, they are given by $\Psi_d=\left[
\begin{array}{c}
           d_\uparrow \\
           d_\downarrow^\dagger \\
\end{array}
\right]$ and $\Psi_k=\left[
\begin{array}{c}
           f_{k\uparrow} \\
           f^\dag_{-k\downarrow} \\
         \end{array}
\right]$.
With the action $S$, we can express the partition function as a path integral, i.e., $Z={\rm Tr}[e^{-\beta H}]=\int\mathcal{D}\Psi_k\mathcal{D}\Psi_k^\dagger e^{-S}$, in which the measure $\mathcal{D}\Psi_k$ denotes all the possible integral paths. After integrating out the fermion field $\Psi_k,\Psi^\dag_k$ with a Gaussian integral, the partition function will become a ``generating functional"
\begin{eqnarray}
Z[\Psi_d,\Psi_d^\dag]&=\det[\partial_\tau+\mathcal{H}_{d0}]\exp\left[\int d\tau d\tau'\sum_k\Psi_d^\dag(\tau) {\cal V}_{kS}\textbf{G}(k,\tau-\tau'){\cal V}^\dag_{kS}\Psi_d(\tau')\right].
\end{eqnarray}
$\textbf{G}(k,\tau-\tau')=-\langle T_\tau\Psi_k(\tau)\otimes\Psi_k^\dag(\tau')\rangle$ is defined in the $s$-wave superconductor. It obeys the Fourier expansion $\textbf{G}(k,\tau)={1\over\beta}\sum_n e^{-i\omega_n\tau}\textbf{G}(k,i\omega_n)$ with
$\textbf{G}(k,i\omega_n)=[\det(i\omega_n-\mathcal{H}_s)]^{-1}\left[
\begin{array}{cc}
 i \omega_n +\xi _k & \Delta_se^{i\theta}   \\
  \Delta_se^{-i\theta}   & i \omega_n -\xi _k \\
\end{array}
\right]$. This allows us to write out the effective expression of the action in the Fourier space directly, i.e.,
\begin{eqnarray}
S_\text{eff}&&=-\sum_{i\omega_n}\text{Tr}\ln G^{-1}_{d0}(i\omega_n)-\frac{t_S^2}{\beta}\sum_{k,i\omega_n}\Psi_d^\dagger \textbf{G}(k,i\omega_n)\Psi_d\notag\\
&&=-\sum_{i\omega_n}\ln\left[\omega_n^2+\varepsilon_0^2\right]-\frac{t_S^2}{\beta}\sum_k\Psi_d^\dagger
[\frac{i\omega_n\sigma_0+\xi_k\sigma_z+\Delta_s\sigma_x}{\omega_n^2+\xi_k^2+\Delta_s^2}]\Psi_d.
\end{eqnarray}
With the help of the expression of $S_\text{eff}$, it is not difficult for us to get the perturbative Hamiltonian of the $s$-wave superconductor on the QD
\begin{eqnarray}
H'\simeq-\frac{t_S^2}{N}\sum_k\Psi_d^\dagger[\frac{\xi_k}{\xi_k^2+\Delta_s^2}\sigma_z +\frac{\Delta_s} {\xi_k^2+\Delta_s^2}\sigma_x]\Psi_d.
\end{eqnarray}
Via a unitary transformation, the system
Hamiltonian can be expressed as the following form
\begin{eqnarray}
H_{\rm eff}=\varepsilon_0\sum_\sigma(n_\sigma-{1\over2})+\tilde{\Delta}_s(e^{i\theta}d_\uparrow^\dagger
d_\downarrow^\dag+e^{-i\theta}d_\downarrow
d_\uparrow)+U(n_\uparrow-{1\over2})(n_\downarrow-{1\over2})+t_T\sum_\sigma(d_\sigma-d_\sigma^\dag)\gamma_{0\sigma},
\end{eqnarray}
with $\tilde{\Delta}_s={t_S^2\over N}\sum_k{\Delta_s
\over\xi_k^2+\Delta_s^2}$. Such a result indicates that the weak Andreev
reflection between the $s$-wave superconductor and QD induces a weak
$s$-wave pairing potential on the QD, which is exactly the so-called the
proximity effect.
\par
For the sake of diagonalizing such a Hamiltonian, we need to introduce local Majorana
operators $\eta_{is}$ through
$d_\sigma=\frac{1}{\sqrt{2}}(\eta_{1\sigma}+i\eta_{2\sigma})$ and
$d_\sigma^\dag=\frac{1}{\sqrt{2}}(\eta_{1\sigma}-i\eta_{2\sigma})$.
And then, by defining Dirac fermionic operators
$a=\frac{1}{\sqrt{2}}(\eta_{1\uparrow}+i\eta_{1\downarrow})$,
$b=\frac{i}{\sqrt{2}}(\eta_{2\uparrow}+i\eta_{2\downarrow})$, and
$\tilde{f}=\frac{1}{\sqrt{2}}(\gamma_{0\uparrow}+i\gamma_{0\downarrow})$,
we can obtain a new expression of $H_{\rm eff}$, i.e.,
\begin{eqnarray}
H_{\rm eff}&=&\tilde{\Delta}_s\sin\theta(a^\dag a-b^\dag
b)+(\varepsilon_0+i\tilde{\Delta}_s\cos\theta)a^\dag
b+(\varepsilon_0-i\tilde{\Delta}_s\cos\theta)b^\dag a\notag\\
&&-U(n_a-{1\over2})(n_b-{1\over2})-\sqrt{2}t_T(b^\dag
\tilde{f}+h.c.).\label{Dirac0}
\end{eqnarray}
\par
According to Eq.(\ref{Dirac0}), the Bogoliubov¨Cde Gennes equation $H_{\rm eff}\Psi=E\Psi$ can be
built, and then the eigenvalues of $H_{\rm eff}$ can be worked out. On the
basis of
$\{|000\rangle,|001\rangle,|010\rangle,|100\rangle,|110\rangle,|101\rangle,|011\rangle,|111\rangle\}$,
the matrix form of $H$ can be obtained
($|n_an_bn_f\rangle=|n_a\rangle |n_b\rangle|n_{\tilde{f}}\rangle$,
where $n_a=a^\dag a$, $n_b=b^\dag b$, and
$n_{\tilde{f}}=\tilde{f}^\dag \tilde{f}$). Note that in the
TS-existed system, only the parity of the average particle
occupation number is the good quantum number, thus the matrix form
of $H_{\rm eff}$ should be given according to fermion parity (FP). In the case
of odd FP,
$\Psi_o=c_1|001\rangle+c_2|010\rangle+c_3|100\rangle+c_4|111\rangle$,
and then
\begin{eqnarray}
 H^{(o)}_{\rm eff}=\left[
\begin{array}{cccc}
 -{U\over4} & -\sqrt{2}t_T &  0 & 0 \\
 -\sqrt{2}t_T & -\tilde{\Delta}_s\sin\theta+{U\over4} & \varepsilon_0-i\tilde{\Delta}_s\cos\theta & 0 \\
0 & \varepsilon_0+i\tilde{\Delta}_s\cos\theta & \tilde{\Delta}_s\sin \theta+{U\over4} & 0\\
 0&0  & 0 &-{U\over4}
\end{array}
\right];
\end{eqnarray}
In the case of even FP, $\Psi_e=c_1|000\rangle+
c_2|011\rangle+c_3|101\rangle+c_4|110\rangle$, and
\begin{eqnarray}
H^{(e)}_{\rm eff}=\left[
\begin{array}{cccc}
 -{U\over4} & 0 &  0 & 0 \\
 0 & -\tilde{\Delta}_s\sin\theta+{U\over4} & \varepsilon_0-i\tilde{\Delta}_s\cos\theta & 0 \\
 0 & \varepsilon_0+i\tilde{\Delta}_s\cos\theta &  \tilde{\Delta}_s\sin \theta+{U\over4} & -\sqrt{2}t_T\\
 0&0  & -\sqrt{2}t_T &-{U\over4}
\end{array}
\right].
\end{eqnarray}
It is easy to find that $H^{(o)}_{\rm eff}(\theta)=H^{(e)}_{\rm eff}(\theta+\pi)$. Thus,
the Josephson effect can be clarified by only analyzing 
the current oscillation result in one FP.

\subsection{Diagonalization of $H$ in Case II} \label{caseII}
In Case II where $t_T\approx t_S$, $H$ is difficult to diagonalize
due to the presence of continuum state in the $s$-wave
superconductor. However, according to the previous works, the zero
band-width approximation is feasible to solve the Josephson effect
contributed by the $s$-wave superconductor.\cite{ZBA} Within such an
approximation, the Hamiltonian can be simplified as
\begin{eqnarray}
H_{\rm eff}&=&\varepsilon_0\sum_\sigma(n_\sigma-{1\over2})+U(n_\uparrow-{1\over2})(n_\downarrow-{1\over2})
+\sum_{\sigma}\xi
f^\dag_{\sigma}f_{\sigma}+\Delta_s(e^{i\theta}f^\dag_{\uparrow}f^\dag_{\downarrow}
+e^{-i\theta}f_{\downarrow}f_{\uparrow})\notag\\
&&+t_T\sum_\sigma(d_\sigma-d_\sigma^\dag)\gamma_{0\sigma}-\sum_{\sigma}t_{S}f_{\sigma}^\dag
d_\sigma+\text{h.c.}.
\end{eqnarray}
\par
By defining
$f_\sigma=\frac{1}{\sqrt{2}}(\zeta_{1\sigma}+i\zeta_{2\sigma})$ and
$f_\sigma^\dag=\frac{1}{\sqrt{2}}(\zeta_{1\sigma}-i\zeta_{2\sigma})$
with
$\alpha=\frac{1}{\sqrt{2}}(\zeta_{1\uparrow}+i\zeta_{1\downarrow})$
and
$\beta=\frac{i}{\sqrt{2}}(\zeta_{2\uparrow}+i\zeta_{2\downarrow})$,
we get the Hamiltonian in the spinless-fermion representation
\begin{eqnarray}
H_{\rm eff}&=&\varepsilon_0(a^\dag b+b^\dag
a)-U(n_a-{1\over2})(n_b-{1\over2})\notag\\
&&+\Delta_s\sin\theta(\alpha^\dag\alpha-\beta^\dag\beta)+(\xi+i\Delta_s\cos\theta)\alpha^\dag
\beta+(\xi-i\Delta_s\cos\theta)\beta^\dag \alpha\notag\\
&&-\sqrt{2}t_T(b^\dag \tilde{f}+\tilde{f}^\dag
b)-t_S(b^\dag\alpha+\alpha^\dag b+a^\dag\beta+\beta^\dag a).\label{Dirac}
\end{eqnarray}
On the basis of odd FP, the matrix of $H^{(o)}_{\rm eff}$ can be
expressed as
\begin{eqnarray}
 H^{(o)}_{\rm eff}=\left[
\begin{array}{ccc}
h^{(o)}_{I}&0&0\\
0 & h^{(o)}_{III} & 0 \\
0&0  &-{U\over4}
\end{array}
\right],
\end{eqnarray}
where
\begin{eqnarray}
h^{(o)}_{I}=\left[
\begin{array}{ccccccc}
 {U\over4} & \varepsilon_0 &  0 & -t_S &0 \\
\varepsilon_0 & {U\over4} & -t_S&0 &-\sqrt{2}t_T\\
0 & -t_S & \Delta_s\sin\theta-{U\over4} &
\xi+i\Delta_s\cos\theta &0\\
-t_S&0  & \xi-i\Delta_s\cos\theta &-\Delta_s\sin\theta-{U\over4} &0\\
0&-\sqrt{2}t_T & 0 & 0&-{U\over4}
\end{array}
\right]\notag
\end{eqnarray}
and
\begin{eqnarray}
h^{(o)}_{III}=\left[
\begin{array}{cccccccccc}
  E_1& t_{N} &  0 & 0 &0 & \sqrt{2}t_T & -t_S &  0 & 0 &0 \\
  t^*_{N}& E_2 &  0 & -t_S &0 & 0 & 0 &  0 & 0 &0 \\
   0 & 0 & -{U\over4} & 0 &0 & -t_S & 0 &  0 & t_S &0 \\
    0 & -t_S & 0& {U\over4} & 0 &0 & \varepsilon_0 & 0 &  0 &0\\
     0 & 0 &  0 & 0 & -E_1 & t^*_{N}& 0 &  0 & \varepsilon_0 &0 \\
      \sqrt{2}t_T & 0 &  -t_S & 0 &t_{N} & -E_2& 0  &  \varepsilon_0 & 0 &t_S \\
       -t_S & 0 &  0 & \varepsilon_0 &0 & 0 & {U\over4} &  0 & 0 &-\sqrt{2}t_T \\
        0 & 0 &  0 & 0 &0 & \varepsilon_0 & 0 &  -E_2 & t_{N} &0 \\
         0 & 0 &  t_S & 0 &\varepsilon_0 & 0 & 0 &  t^*_{N} & -E_1 &-t_S\\
          0 & 0 &  0 & 0 &0 & t_S & -\sqrt{2}t_T &  0 & -t_S &-{U\over4}
\end{array}
\right]\notag
\end{eqnarray}
with $E_1=\Delta_s\sin\theta-{U\over4}$, $E_2=-\Delta_s\sin\theta-{U\over4}$, and $t_{N}=\xi+i\Delta_s\cos\theta$.
On the basis of even FP, the matrix of $H^{(e)}_{\rm eff}$ can be
given by
\begin{eqnarray}
 H^{(e)}_{\rm eff}=\left[
\begin{array}{ccc}
-{U\over4}&0&0\\
0 & h^{(e)}_{IV} & 0 \\
0&0  &h^{(e)}_{II}
\end{array}
\right],
\end{eqnarray}
where
\begin{eqnarray}
h^{(e)}_{IV}=\left[
\begin{array}{ccccccc}
 {U\over4} & \varepsilon_0 &  0 & -t_S &0 \\
\varepsilon_0 & {U\over4} & -t_S&0 &-\sqrt{2}t_T\\
0 & -t_S & -\Delta_s\sin\theta-{U\over4} &
\xi-i\Delta_s\cos\theta &0\\
-t_S&0  & \xi+i\Delta_s\cos\theta &\Delta_s\sin\theta-{U\over4} &0\\
0&-\sqrt{2}t_T & 0 & 0&-{U\over4}
\end{array}
\right]\notag
\end{eqnarray}
and
\begin{eqnarray}
h^{(e)}_{II}=\left[
\begin{array}{cccccccccc}
  E_2& t^*_{N} &  0 & 0 &0 & \sqrt{2}t_T & -t_S &  0 & 0 &0 \\
  t_{N}& E_1 &  0 & -t_S &0 & 0 & 0 &  0 & 0 &0 \\
   0 & 0 & -{U\over4} & 0 &0 & -t_S & 0 &  0 & t_S &0 \\
    0 & -t_S & 0& {U\over4} & 0 &0 & \varepsilon_0 & 0 &  0 &0\\
     0 & 0 &  0 & 0 & -E_2 & t_{N}& 0 &  0 & \varepsilon_0 &0 \\
      \sqrt{2}t_T & 0 &  -t_S & 0 &t^*_{N} & -E_1& 0  &  \varepsilon_0 & 0 &t_S \\
       -t_S & 0 &  0 & \varepsilon_0 &0 & 0 & {U\over4} &  0 & 0 &-\sqrt{2}t_T \\
        0 & 0 &  0 & 0 &0 & \varepsilon_0 & 0 &  -E_1 & t^*_{N} &0 \\
         0 & 0 &  t_S & 0 &\varepsilon_0 & 0 & 0 &  t_{N} & -E_2 &-t_S\\
          0 & 0 &  0 & 0 &0 & t_S & -\sqrt{2}t_T &  0 & -t_S &-{U\over4}
\end{array}
\right].\notag
\end{eqnarray}
Therefore, the different-FP matrix forms of $H_{\rm eff}$ obeys the relationship $H^{(o)}_{\rm eff}(\theta)=H^{(e)}_{\rm eff}(\theta+\pi)$, similar to the result in Case I.

\subsection{Diagonalization of $H$ in Case III}
In Case III, we turn to the discussion about the diagonalization of $H$ when $t_T\ll t_S$. In such a
case, the QD will dip in the $s$-wave superconductor, leading to the
formation of a composite $s$-wave superconductor. Consequently, the
considered structure will be transformed into a junction in which the
Majorana doublet couples to a $s$-wave superconductor directly. Its
Hamiltonian can thus be written as
\begin{eqnarray}
H_{\rm eff}=\sum_{k\sigma}
W_{k}(F_{k\sigma}-F_{k\sigma}^\dag)\gamma_{0\sigma}
+\sum_{k\sigma}\xi_kF^\dag_{k\sigma}F_{k\sigma}+\Delta_s\sum_k(e^{i\theta}F^\dag_{k\uparrow}F^\dag_{-k\downarrow}
+e^{-i\theta}F_{-k\downarrow}F_{k\uparrow}).\label{caseIII}
\end{eqnarray}
In Eq.(\ref{caseIII}), $F_{k\sigma}$ originates from the unitary
transformation that $d_\sigma=\sum_k\nu_kF_{k\sigma}$ and
$f_{k\sigma}=\sum_{k'}\eta_{kk'}F_{k'\sigma}$, and $W_k=\nu_k t_T$
denotes the coupling between the Majorana doublet and the $s$-wave
superconductor. It is easy to prove that in composite $s$-wave superconductor couples weakly to the Majorana doublet [See the appendix]. As a result, the composite $s$-wave superconductor
can be considered as perturbation. With respect to the Hamiltonian in Eq.(\ref{caseIII}), the action
can be written as
\begin{eqnarray}
S&=\int d\tau\ \sum_k\left\{\tilde{\Psi}^\dag_k[\partial_\tau+{\cal H}_s]\tilde{\Psi}_k+[{\cal W}(k)\Gamma_0^\dagger\tilde{\Psi}_k+{\cal W}^\dag(k)\tilde{\Psi}_k^\dag\Gamma_0]
+\Gamma_0^\dagger\partial_\tau\Gamma_0\right\},
\end{eqnarray}
in which ${\cal H}_s=\left[
\begin{array}{cc}
 \xi _k & \Delta_se^{i\theta}  \\
  \Delta_se^{-i\theta}  & -\xi _k \\
\end{array}
\right]$ and ${\cal W}(k)=W_k\sigma_z$ with $\tilde{\Psi}_k=\left[
\begin{array}{c}
           F_{k\uparrow} \\
           F^\dag_{-k\downarrow} \\
         \end{array}
\right]$ and $\Gamma_0=\left[
\begin{array}{c}
           \gamma_{0\uparrow} \\
           \gamma_{0\downarrow}\\
\end{array}
\right]$.
Similar to the derivation process in Case I, we express the partition function as a path integral \begin{eqnarray}
Z=\int\mathcal{D}\tilde{\Psi}_k\mathcal{D}\tilde{\Psi}_k^\dagger e^{-S}
=\int\mathcal{D}\tilde{\Psi}_k\mathcal{D}\tilde{\Psi}_k^\dagger\exp\left[-\int d\tau\;\sum_k\left\{\tilde{\Psi}^\dag_k[\partial_\tau+\mathcal{H}_s(k)]\tilde{\Psi}_k+({\cal W}\Gamma_0^\dagger\tilde{\Psi}_k+{\cal W}^\dag\tilde{\Psi}^\dagger_k\Gamma_0)\right\}\right].\notag
\end{eqnarray}
Integrating out the fermion field $\tilde{\Psi}_k,\tilde{\Psi}^\dag_k$ with a Gaussian integral, we simplify the partition function as
\begin{eqnarray}
Z[\Gamma_0,\Gamma_0^\dagger]&=&\det[\partial_\tau]\exp\left[\int d\tau d\tau'\sum_k\Gamma_0^\dag(\tau) {\cal W}\tilde{\textbf{G}}(k,\tau-\tau'){\cal W}^\dag\Gamma_0(\tau')\right]\notag\\
&=&\det[\partial_\tau]\exp\left[\int d\tau d\tau'\Gamma_0^\dag(\tau) \tilde{t}_T\textbf{G}_d(\tau-\tau')\tilde{t}_T^\dag\Gamma_0(\tau')\right],
\end{eqnarray}
where $\tilde{\textbf{G}}(k,\tau-\tau')=-\langle T_\tau\tilde{\Psi}_k(\tau)\otimes\tilde{\Psi}_k^\dag(\tau')\rangle$, $\textbf{G}_d(\tau-\tau')=-\langle T_\tau\Psi_d(\tau)\otimes\Psi_d^\dag(\tau')\rangle$, and
$\tilde{t}_T=t_T\sigma_z$.
\par
Since $\textbf{G}_d(\tau)$ obeys the relationship that $\textbf{G}_d(\tau)={1\over\beta}\sum_n e^{-i\omega_n\tau}\textbf{G}_d(i\omega_n)$, in the Fourier space the effective action can be expressed as
$S_\text{eff}=-\sum_{i\omega_n}\text{Tr}\ln [\omega_n^2]-\frac{1}{\beta}\sum_{i\omega_n}\Gamma_0^\dag(i\omega_n)\tilde{t}_T\textbf{G}_d(k,i\omega_n)\tilde{t}_T^\dag\Gamma_0(i\omega_n)\notag$.
Accordingly, the Josephson Hamiltonian in Case III can be given by
$H_\text{eff}(\theta)=-\frac{t_T^2}{\beta}\sum_{i\omega_n}\left[2i {\rm Im}G_{d,he}(i\omega_n)\right]\gamma_{0\uparrow}\gamma_{0\downarrow}$.
At the zero-frequency limit, the approximated form of
$H_{\rm eff}$ can be written as $H_{\rm
eff}=J(i\gamma_{0\uparrow}\gamma_{0\downarrow})\sin\theta$ with
\begin{equation}
J=-it_T^2\langle d^\dag_\uparrow d^\dag_\downarrow\rangle\approx
\Delta_st_T^2/t_S^2.
\end{equation}
By defining
$\tilde{f}={1\over\sqrt{2}}(\gamma_{0\uparrow}+i\gamma_{0\downarrow})$,
we obtain the result that $H_{\rm
eff}=J(n_{\tilde{f}}-{1\over2})\sin\theta$, i.e., $H^{(o/e)}_{\rm
eff}=(+/-){J\over2}\sin\theta$. Therefore, the
Josephson current can be directly written as
$I_J^{(o/e)}=[{e\over \hbar}](+/-)J\cos\theta$. Surely, such a result is consistent with that in Ref.\cite{Qixl}.

\begin{figure}[htb]
\begin{center}
\scalebox{0.4}{\includegraphics{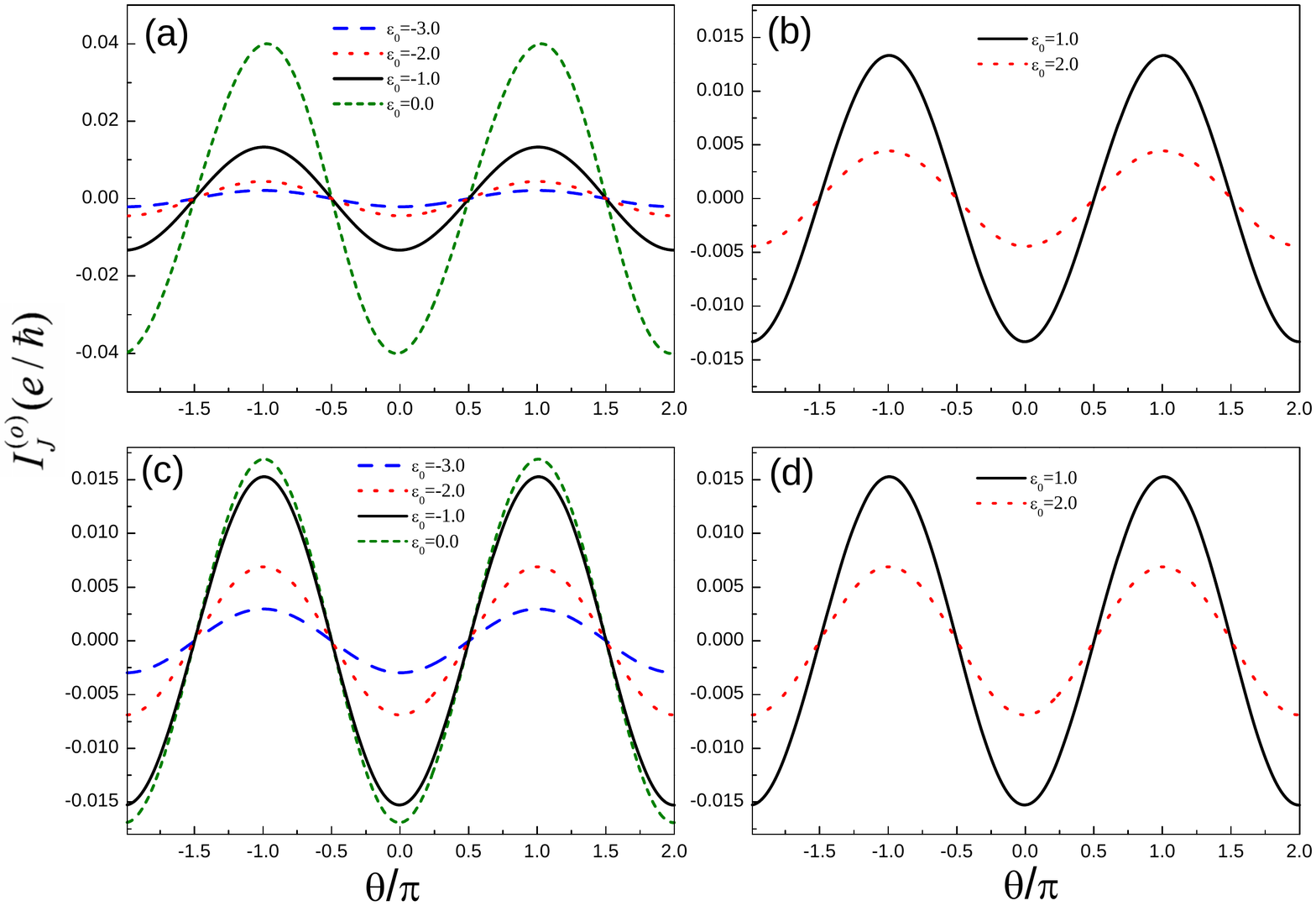}} \caption{Spectra of odd-FP Josephson
current in Case I of $U=0$ and $U=2.0$, respectively. The coupling strength between the QD and superconductors are taken to be $t_T=0.5$ and $\tilde{\Delta}_s=0.04$. (a)-(b) $U=0$; (c)-(d) $U=2.0$.} \label{Case1}
\end{center}
\end{figure}

\section{Numerical results and discussions}
With the theory in the above section, we proceed to calculate the Josephson current in various cases. As a typical case, the system temperature is taken to be zero. Besides, we take $\Delta_s=1.0$ to be the energy unit in this junction.
\begin{figure}[htb]
\begin{center}
\scalebox{0.4}{\includegraphics{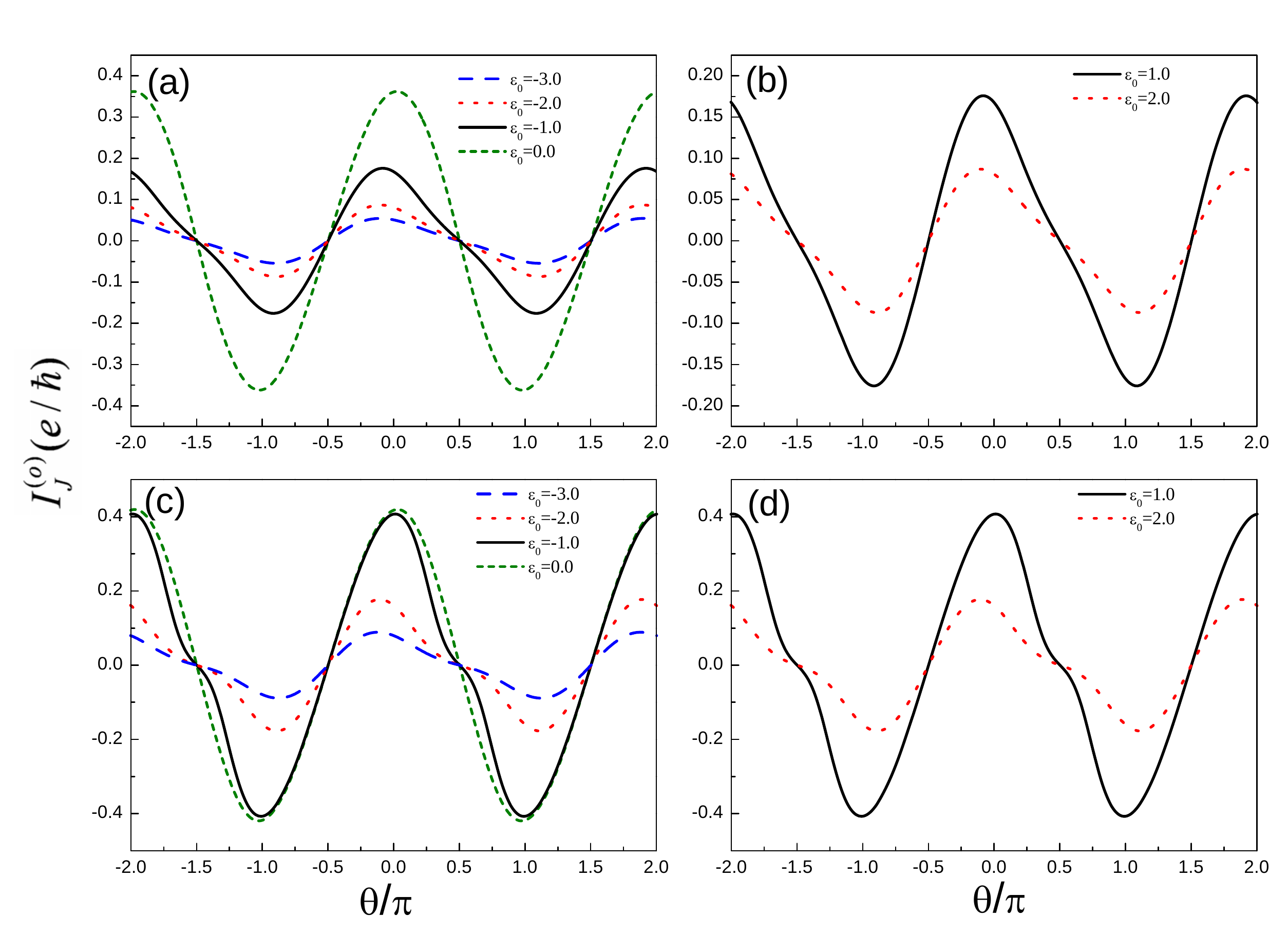}} \caption{Odd-FP current in Case II of $U=0$ and $U=2.0$. In (a)-(b) $U=0$, and $U=2.0$ in (c)-(d). Relevant parameters are chosen as $t_T=t_S=0.5$.}
\label{Case2}
\end{center}
\end{figure}
\par
In Fig.\ref{Case1}, we investigate the odd-FP Josephson current in Case I, and plot its spectra as a function of Josephson phase difference. As for the QD level and intradot Coulomb strength, we change $\varepsilon_0$ from $-3.0$ to $2.0$ in which $U$ is taken to be $0.0$ and $2.0$, respectively. In this figure, we find that despite the change of $\varepsilon_0$ and $U$, the leading oscillation property of the Josephson current is fixed. Namely, it reaches the maximum at the point $\theta=n\pi$ with its profile as $I^{(o)}_J\sim-\cos\theta$. However, the roles of QD level and Coulomb strength can also be clearly observed. With the departure of the QD level from energy zero point, the current amplitude will be suppressed gradually. Such a result is relatively apparent in Fig.\ref{Case1}(a)-(b) where reflects the case of the zero Coulomb interaction. This can be explained as the weakness of the quantum coherence when the QD level is away from energy zero point. The effect of Coulomb interaction is notable in the region of $\varepsilon_0$, where the QD level is occupied. It can be found that Coulomb interaction suppresses the current amplitude as well. This should be attributed to the destruction of the quantum coherence induced by the QD-level splitting ($\varepsilon_0\to \varepsilon_0 $ and $\varepsilon_0+U$) in the presence of Coulomb interaction.

\begin{figure}[htb]
\begin{center}
\scalebox{0.4}{\includegraphics{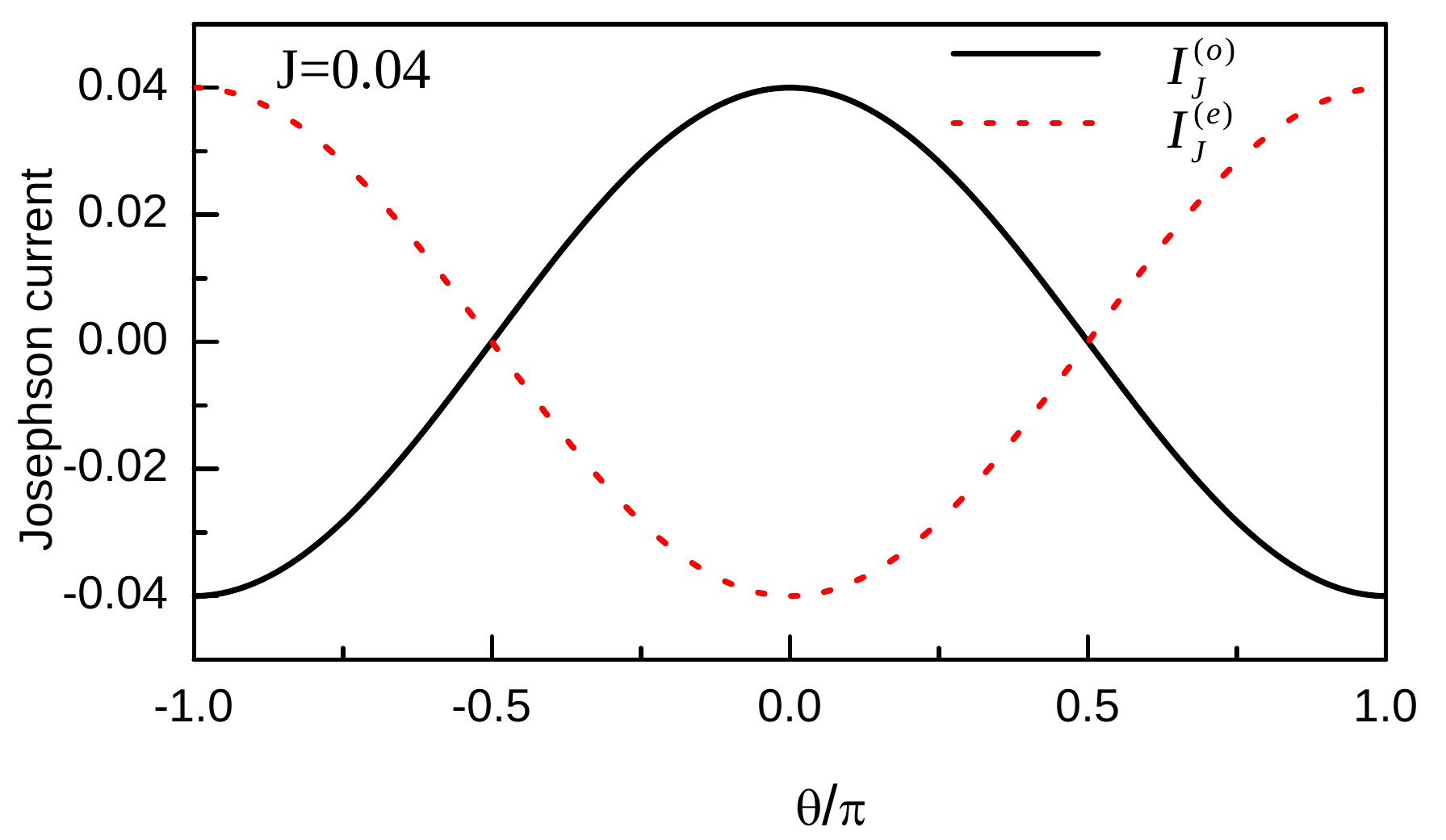}} \caption{Josephson currents in Case III. The relevant parameters are taken to be $t_T=0.1$ and $t_S=0.5$. In the case of $\Delta_s=1.0$, the current amplitude $J$ will be equal to be $0.04$.}
\label{Case3}
\end{center}
\end{figure}
\par

\begin{figure}[htb]
\begin{center}
\scalebox{0.4}{\includegraphics{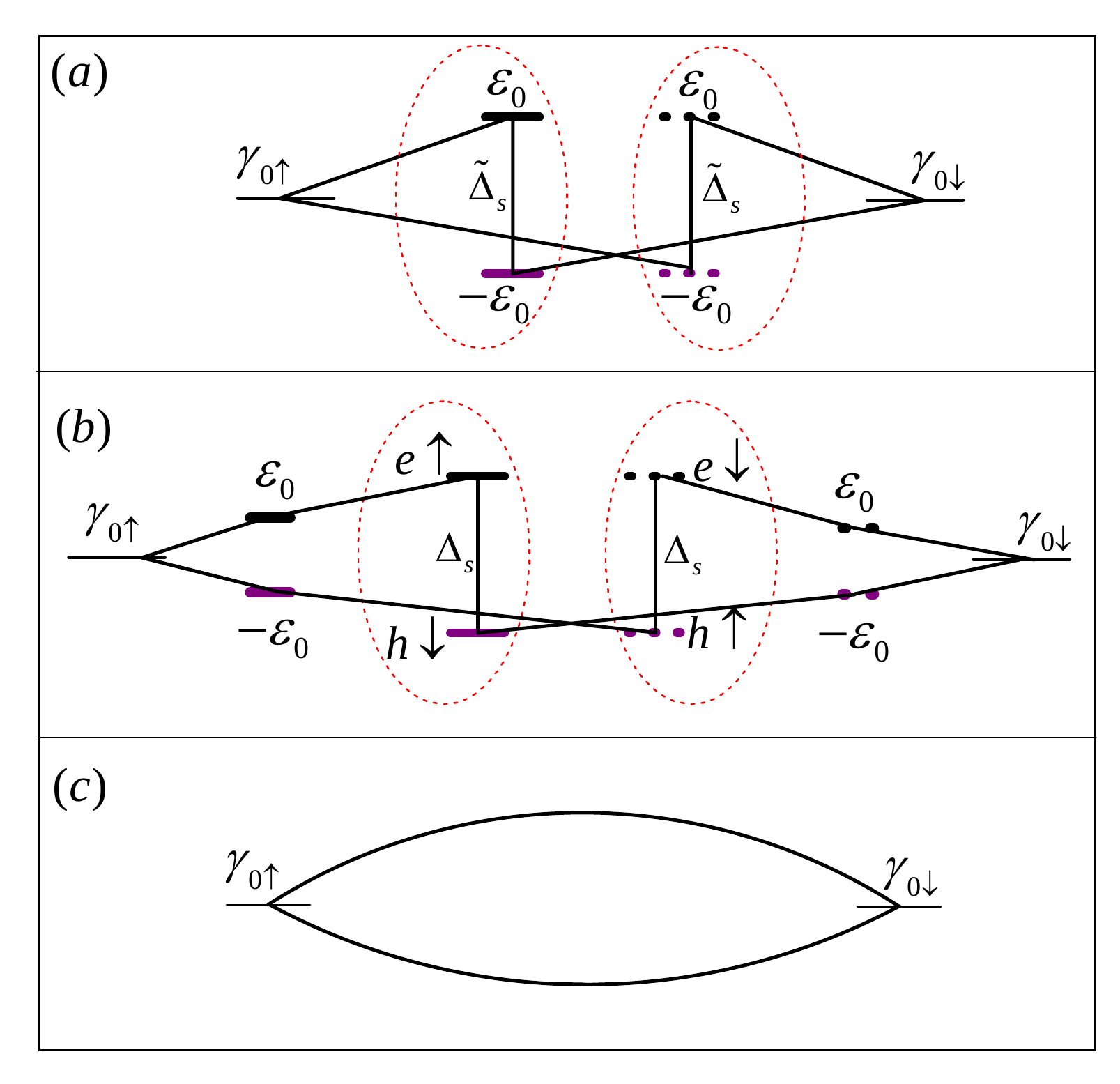}}
\caption{(a)-(b) Geometries of the Josephson junction in Case I, Case II, and Case III in the Nambu representation, respectively.}
\label{Nambu}
\end{center}
\end{figure}

According to the discussion about Case II in the second part of Sec.II, when $t_S$ gets close to $t_T$, the $s$-wave superconductor should not be viewed as perturbation. It is natural to think that the Josephson current in such a case will show new properties. Next, we would like to increase the coupling strength between the QD and $s$-wave superconductor to discuss the change of Josephson effect. The numerical results are shown in Fig.\ref{Case2} where the $t_T=t_S=0.5$. In this figure, we see that in such a case, the current properties are completely different from those in Case I. Firstly, the current amplitude is efficiently enhanced by the increase of $t_S$. Secondly, the current direction is completely reversed and its profile is deviated from $I_J^{(o)}\sim \cos\theta$.
If the coupling between the QD and $s$-wave superconductor further increases, the QD will be submerged in the $s$-wave superconductor. Consider the extreme case of the weak-coupling limit (i.e., Case III), the perturbation method can also be employed to evaluate the Josephson current, as displayed in the third part in Sec.II. It clearly shows that in such a case, the Majorana doublet couples weakly to the composite $s$-wave superconductor. Consequently, $I_J^{(o/e)}$ follow the relationship that $I_J^{(o/e)}=\pm J\cos\theta$ and their properties are clearly shown [See the results in Fig.\ref{Case3}].
\par
In view of the current results in Case I, Case II, and Case III, one can observe that at the case of $t_T\gg t_S$ (i.e., Case I), the current oscillation manner is opposite to that in the other two cases.
In order to clarify the change of Josephson current from Case I to Case III, we present the geometries of these three cases in the Nambu representation, as shown in Fig.\ref{Nambu}. In Fig.\ref{Nambu}(a)-(c), we notice that the finite coupling between the two MBSs of Majorana doublet, despite the direct or indirect coupling, give rise to the occurrence of anomalous Josephson effect. On the other hand, the coupling strength between the QD and $s$-wave superconductor plays the role in altering the inter-MBS coupling property, leading to the change of the current oscillation manner. In the case of $t_T\ll t_S$, the MBSs couple directly to each other with a constant coupling parameter. In such case, the current direction is only dependent on the FP of the Majorana doublet with $I_J=[{e\over \hbar}]{\cal P}J\cos\theta$. In the other case where $t_T\gg t_S$, the coupling between the QD and Majorana doublet induces the indirect inter-MBS coupling, as shown in Fig.\ref{Nambu}(a). With respect to the inter-MBS coupling in these two cases, we find that in the former case, the MBSs couple to each other via a nonresonant Andreev reflection process, whereas in the latter case, one bound state is involved in the Andreev reflection process. It is well known that the quasi-particle phase will undergo a $\pi$-phase shift due to the presence of one bound state in the Andreev reflection process. Accordingly, for the case of identical FP, the current oscillations in Case I and Case III are opposite to each other. By the same token, we can easily see
that in Case I and Case II, the oscillation manners of the Josephson current are opposite to each other, since an additional bound state is presented in the Andreev reflection process in Case II. Up to now, we have known the reason that in the considered junction, the current in the case of $t_T\gg t_S$ is different from that in the other cases. Also, note that in such a structure, the role of the $s$-wave superconductor is to provide a channel for the coupling between the MBSs in the Kramers doublet and the QD is to change the channel property. For this reason, the change of $\varepsilon_0$ and $U$ cannot induce any phase transition behaviors for the Josephson effect.  
\par

\section{Summary}
In summary, in this work we have discussed the Josephson effects in the junction formed by a one dimensional \emph{DIII}-class TS
and a $s$-wave superconductor, by embedding a
QD in such a junction. Via considering three QD-superconductor coupling
manners, we have presented a comprehensive analysis about the Josephson effect in this system. As a consequence, it has been found that the Josephson current oscillates in $2\pi$ period. However, the presence of Majorana doublet in the \emph{DIII}-class TS renders the Josephson
current finite in the case of zero phase difference between the superconductors. With respect to the current direction, it is not only related to the FP of this junction but also depends on the coupling strength between the QD and $s$-wave superconductor. When the coupling between the QD and $s$-wave superconductor decreases to its weak limit, the direction of the Josephson current will have an opportunity to reverse. Via analyzing the particle motion in this structure, the reason for such a result has been clarified, namely, the QD-superconductor coupling manner can alter the property of the Andreev reflection between the MBSs in the Majorana doublet. It can be believed that this work be helpful for understanding the transport properties of the \emph{DIII}-class TS.


\section*{Appendix}
According to the Hamiltonian in Eq.(\ref{caseIII}), the coupling strength between the Majorana doublet and the $s$-wave superconductor, i.e.,
$\tilde{\Gamma}_\sigma=\pi\sum_k|W_k|^2\rho_\sigma$, where $\rho_\sigma$ is the density of state in the $s$-wave
superconductor. Via a straightforward derivation, one can get the result that $\tilde{\Gamma}_\sigma=-t_T^2{\rm Im}G^r_{d\sigma}$.
$G^r_{d\sigma}$, defined by $G^r_{d\sigma}(t)=-i\theta(t)\langle \{d_{\sigma}(t), d_\sigma^\dag\}\rangle$, is a retarded Green function of one QD coupled to a $s$-wave superconductor. By means of the nonequilibrium Green function technique, the matrix of the retarded
Green function $\textbf{G}^r_d$ can be obtained, i.e.,
\begin{small}
\begin{eqnarray}
[\textbf{G}^r_d]^{-1}&=&\left[
\begin{array}{cccc}
 (\omega-\varepsilon_0)S_{e\uparrow}+i\Gamma_0\rho_0& 0 & 0 & -i{\Delta_s\over \omega}\Gamma_0\rho_0\\
 0 & (\omega-\varepsilon_0)S_{e\downarrow}+i\Gamma_0\rho_0& -i{\Delta_s\over \omega}\Gamma_0\rho_0& 0 \\
 0 & -i{\Delta_s\over \omega}\Gamma_0\rho_0& (\omega+\varepsilon_0)S_{h\uparrow}+i\Gamma_0\rho_0& 0 \\
 -i{\Delta_s\over \omega}\Gamma_0\rho_0 & 0 & 0 & (\omega +\varepsilon_0)S_{h\downarrow}+i\Gamma_0\rho_0\\
\end{array}
\right],
\end{eqnarray}
\end{small}
where $\Gamma_0=\pi\sum_k|V_{kS}|^2\delta(\omega-\xi_k)$ and
$\rho_0={|\omega|\over\sqrt{\omega^2-\Delta_s^2}}$. $S_{e(h)\sigma}={\omega\pm\varepsilon_0\pm
U\over\omega\pm\varepsilon_0\pm U\mp U\langle
n_{\bar\sigma}\rangle}$ is the effect of the electron interaction within the Hubbard-I approximation and $\langle n_{\bar\sigma}\rangle$ is the
average electron occupation number. In the absence of magnetic factors, such a system is spin-degenerated, hence $\textbf{G}^r_d$ can be simplified to be $2\times2$ matrix, i.e.,
\begin{eqnarray}
\textbf{G}^r_d=\left[
\begin{array}{cccc}
 (\omega-\varepsilon_0)S_{e}+i\Gamma_0\rho_0& -i{\Delta_s\over \omega}\Gamma_0\rho_0\\
 -i{\Delta_s\over \omega}\Gamma_0\rho_0 & (\omega +\varepsilon_0)S_{h}+i\Gamma_0\rho_0\\
\end{array}
\right]^{-1}.
\end{eqnarray}
In the limit of strong QD-superconductor coupling, the influence of $\omega$ and
$\varepsilon_0$ will be submerged, and $G^r_{d\sigma}$ can be
approximated as $-i{\rho_0/\Gamma_0}$, thus
$\tilde{\Gamma}_\sigma\approx {t_T^2\rho_0/\Gamma_0}$. With
the increment of $t_S$, $\Gamma_0$ will increase. This surely leads to the decrease of $\tilde{\Gamma}_\sigma$.

\clearpage

\bigskip

\end{document}